\documentclass[pdflatex,sn-chicago]{sn-jnl}

\usepackage{graphicx}%
\usepackage{multirow}%
\usepackage{amsmath,amssymb,amsfonts}%
\usepackage{amsthm}%
\usepackage{mathrsfs}%
\usepackage[title]{appendix}%
\usepackage{xcolor}%
\usepackage{textcomp}%
\usepackage{manyfoot}%
\usepackage{booktabs}%
\usepackage{algorithm}%
\usepackage{algorithmicx}%
\usepackage{algpseudocode}%
\usepackage{listings}%
\usepackage{hyperref}%
\hypersetup{
  colorlinks   = true, 
  urlcolor     = blue, 
  linkcolor    = blue, 
  citecolor   = blue 
}
\usepackage{comment}%
\usepackage{float} 
\usepackage{booktabs} 
\usepackage{caption}   

\theoremstyle{thmstyleone}%
\theoremstyle{thmstylethree}%

\begin{document}

\title[Openness to Immigration]{De facto Openness to Immigration}


\author*[1,2,3]{\fnm{Ljubica} \sur{Nedelkoska}}\email{nedelkoska@csh.ac.at}

\author*[4]{\fnm{Diego} \sur{Martin}}\email{dimartin@hbs.edu}

\author*[1]{\fnm{Alexia} \sur{Lochmann}}\email{Alexia\_Lochmann@hks.harvard.edu}

\author[1]{\fnm{Ricardo} \sur{Hausmann}} \email{ricardo\_hausmann@hks.harvard.edu}

\author[1]{\fnm{Dany} \sur{Bahar}}\email{dany\_bahar@hks.harvard.edu}

\author[1]{\fnm{Muhammed A.} \sur{Yildirim}}\email{muhammed\_yildirim@hks.harvard.edu}

\affil[1]{\orgname{Growth Lab at Harvard University}, \orgaddress{\city{Cambridge}, \state{MA}, \country{USA}}}

\affil*[2]{\orgname{Complexity Science Hub}, \orgaddress{\city{Vienna}, \country{Austria}}}

\affil[3]{\orgname{Central European University}, \orgaddress{\city{Vienna},  \country{Austria}}}

\affil[4]{\orgname{Harvard Business School}, \orgaddress{\city{Boston}, \state{MA}, \country{USA}}}

\abstract{Various factors influence why some countries are more open to immigration than others. Policy is only one of them. We design country-specific measures of openness to immigration that aim to capture \textit{de facto} levels of openness to immigration, 
complementing existing \textit{de jure} measures of immigration, based on enacted immigration laws and policy measures. We estimate these for 148 countries and three years (2000, 2010, and 2020). For a subset of countries, we also distinguish between openness towards tertiary-educated migrants and less than tertiary-educated migrants. Using the measures, we show that most places in the World today are closed to immigration, and a few regions are very open. The World became more open in the first decade of the millennium, an opening mainly driven by the Western World and the Gulf countries. Moreover, we show that other factors equal, countries that increased their openness to immigration, reduced their old-age dependency ratios, and experienced slower real wage growth, arguably a sign of relaxing labor and skill shortages.}



\maketitle

\section{Introduction}\label{sec:intro}

Concerns over immigration were at the forefront of most British voters' minds when they voted to leave the European Union in 2016 \citep{yougovMostImportant}. Since the Brexit vote, the UK has introduced stricter visa rules for immigrants with less than tertiary education and has continued its hostile environment policy. \citep{oxIntegrationPostBrexit}. Yet, according to the Annual Population Survey data of 2024\footnote{As reported by the Migration Observatory at the University of Oxford \citep{oxMigrantsOverview}}, the number of foreign nationals in the UK increased from less than 9.2 million in 2016 to over 10.3 million in 2023. What is more, the new cohorts of migrants were less likely to come from the EU, according to that same survey and the World Bank’s immigration data \citep{worldbank2023}, and were less likely to be tertiary educated. In spite of the UK becoming more closed to less skilled migration \textit{de jure}, it became more open \textit{de facto}, and in spite of it becoming more open to high skilled migration \textit{de jure}, it became less open \textit{de facto}.

The Brexit example shows that immigration policy is one out of many factors shaping places' openness to migration. Barriers to immigration extend beyond policies that restrict physical entry; they also exist in labor laws that limit the hiring of foreign workers \citep{umana2019visas} and in property ownership and business regulations that disadvantage non-citizens \citep{unctad2016}. Professional accreditation requirements and restrictions on certain professions further exclude immigrants from economic opportunities \citep{McHugh2017}. Beyond legal frameworks, informal labor market practices in host countries can create additional, less visible obstacles to integration \citep{dancygier2014immigration}. Hence, migration policy changes do not always have the desired outcomes \citep{theoharides2020unintended}. However, to the best of our knowledge, there exist no measures of countries' \textit{de facto} openness to immigration, and our major contribution is providing such measures. 

We could, of course, look at what is happening to actual migration levels to get a sense of \textit{de facto} openness of countries. However, raw immigration rates are not a proper indicator of openness. Some countries have attracted more immigrants due to their central geographic location, their relative wealth, language proximity, and historical ties with other places, among other factors \citep{cavaille2023immigration, morales2024escaping}. Some countries are in proximity to places that have experienced conflict or natural disasters. Our contribution is based on the insight that parsing such factors of immigration from the actual immigration rates of countries, gives the residual migration, which contains information about how open a country is to immigration, \textit{de facto}. To operationalize this, we use gravity models of migration, where bilateral migration stocks are modeled as a function of geographic distances, language proximity, historical ties, and contiguity between countries, as well as country-level indicators such as income level, population, and land area. The resulting residuals (the difference between the actual and the predicted levels of migration) are the building blocks of our openness measures. Moreover, we may want to distinguish between a country that hosts immigrants from a single or a couple of destinations, other factors equal, and a destination that hosts immigrants from several different countries. Even if the two host the same number of immigrants relative to their population, we may intuitively think of the second one as being \textit{more open}.

We show that diversity-based measures (the number of positive residuals) perform better than scale-based measures (the overall size of the bilateral residuals) in capturing our concept of openness to immigration. An intuitive interpretation of the first is that they measure a country’s openness to people who are geographically, linguistically, and historically different from the dominant host country population. We find that the countries performing best on these measures are the Western Offshoots (USA, Australia, Canada, New Zealand), much of Western Europe, Sweden, and Denmark. The Gulf countries, which due to the extraordinary size of their foreign-born labor markets, perform best on scale-based measures, are moderately diverse \textit{de facto}, and their diversity has been increasing over time. The analysis also reveals that most of the World is closed to immigration. At the global level, between 2000 and 2010, countries, on average, increased their openness to one additional place, while between 2010 and 2020, openness increased more slowly - by less than one place. These averages, however, hide the unequal pattern of opening to immigration, with a handful of countries experiencing vast increases in diversity while others becoming more closed.

The measures are available for three years (2000, 2010, and 2020), and 148 countries. For a subset of countries, we also estimate them by broad skill groups of immigrants - tertiary educated and less than tertiary educated, for the years 2000 and 2010. The measures complement existing ones of migration policy and migrant integration policies (such as DEMIG, POLMIG, MIPEX and IMPALA), such that actual changes in policies can be compared to changes in openness. Moreover, our measures cover most countries in the world with a population of 1.2 million or greater.  The panel can be extended as soon as a new wave of decennial data becomes available. Most importantly, the measures give us a way to gauge the effectiveness of migration policy. Back to the example of Brexit, according to our openness measure, the UK's openness to immigration increased from 33 places in 2000, to 55 in 2010 and to 62 in 2020, while compositionally shifting away from skilled immigration between 2010 and 2020. The intentions of the UK immigration policies have not been reflected in its \textit{de facto} openness immigration.

The measures have high construct validity - they are highly correlated with \textit{de jure} measures of immigration, and with Gallup's global Migrant Acceptance Index (MAI). Moreover, our findings indicate that measures of openness to immigration can predict macro-level demographic and economic outcomes. Specifically, we examine how increased immigration openness affects demographic shifts and real wage changes, the latter serving as a proxy for labor shortages. Our analysis reveals that, holding other factors constant, greater openness to immigration is associated with a lower old-age dependency ratio and slower real wage growth. 
 
\section{Data}\label{data}
Our data comes from four sources. The data on migration stocks comes from the United Nations' global matrices of bilateral migrant stocks for 2000, 2010, and 2020 \citep{UN2024}. The UN use population and housing censuses are a key source of data on the international migrant stock. The information about the share of tertiary and non-tertiary immigrants in the total migrant stock is based on estimates by the World Bank as used in the World Development Report 2023 \citep{worldbank2023}, but we limit its use to the non-imputed part of the dataset, i.e., the part that stems from the countries included in the DIOC-Extended dataset \citep{oecd_dioc_2020}. Most explanatory variables for the Gravity model come from the CEPII Gravity database \citep{conte2022cepii}, and the GDP p.c. PPP, total population and the land area variables come from the World Bank Development Indicators \citep{WorldBank2024}. 

The bilateral migrant stock ($M_{od}$) is measured in terms of numbers of foreign-born people age 15 or older\footnote{\cite{UN2024} primarily use place of birth, as opposed to nationality or citizenship, to identify the immigrant population.}, and for a subset of countries, we can also distinguish between tertiary-educated foreign-born and less-than-tertiary educated foreign-born\footnote{We only include the countries that are either part of DIOC or DIOC-E datasets \citep{oecd_dioc_2020} and which at any point in 2000, 2010, or 2020 had a population of at least 1.2 million. These include 91 destination countries in 2000 and 81 in 2010. They also include 148 countries of origin. Whenever possible in these data, immigrants are identified by their place of birth, but where this information is unavailable, citizenship was used instead.}.

GDP per capita PPP ($GDPpc_{d}$)  provides per capita values for the gross domestic product (GDP) expressed in current international dollars converted by purchasing power parity (PPP) conversion factor. $pop_{d}$ is a mid-year population based on the \emph{de facto} definition of population, which counts all residents regardless of legal status or citizenship. $land_{d}$ is the log of the land area measured in squared kilometers and was originally collected by the Food and Agriculture Organization.    

The definitions of the CEPII-based variables are as follows \citep{CEPII2022}.
$dist_{od}$ is the main city to main city distance between countries. Main cities are identified as in the UN World Urbanization Prospects dataset \citep{UNWUP2024}. 
$contig_{od}$ is a dummy variable that takes a value of one if two countries are contiguous and zero otherwise. Country boundaries are defined as of January 2020 using the ARCGIS’s World Countries (Generalized) dataset \citep{ArcGIS2020}. The dummy variable $comcol_{od}$ equals one if countries share a common colonizer post 1945. The dummy variable $coldepever_{od}$ equals one if a pair was ever in a colonial or dependency relationship (including before 1945). Table \ref{tab:summary_stats} presents the descriptive statistics of the number of migrants, GDP, population, and the variables from the CEPII database.


\begin{table}[h]
  \centering
  \caption{Summary statistics \label{tab:summary_stats}}
    \begin{tabular}{lcccc}
    \midrule
     & \textbf{Std. Dev.} & \textbf{Min} & \textbf{Max} \\
    \midrule
   \multicolumn{2}{l}{Panel A: All immigrants}        &       &       &  \\ \hline
    \multicolumn{1}{l}{Migrants } &                       9,890  &                     116,097  & 0     &                   12,200,000  \\
    \multicolumn{1}{l}{Population } &            48,100,000  &            158,000,000  &            1,230,849  &             1,410,000,000  \\
    \multicolumn{1}{l}{GDP p.c. PPP, destination country} &                    15,912  &                       17,716  &                        433  &                         100,226  \\
    \multicolumn{1}{l}{Distance in km between main cities} &                       7,105  &                         4,192  &                            8  &                           19,819  \\
    \multicolumn{1}{l}{Contiguity} &                         0.03  &                            0.16  & 0     & 1 \\
    \multicolumn{1}{l}{Common official or primary language} &                         0.12  &                            0.33  & 0     & 1 \\
    \multicolumn{1}{l}{Common colonizer post 1945} &                         0.07  &                            0.26  & 0     & 1 \\
    \multicolumn{1}{l}{Ever in colonial relationship} &                         0.01  &                            0.11  & 0     & 1 \\
    \multicolumn{1}{l}{Land area in square km (2020)} &                  877,418  &                 2,075,954  & 718   &                   16,400,000  \\
    \midrule
    \multicolumn{1}{r}{} &       &       &       &  \\
    \multicolumn{2}{l}{Panel B: Sample with level education breakdown}       &       &       &  \\
    \midrule
    \multicolumn{1}{l}{Tertiary migrants } &                       2,991  &                       28,882  & 0     &                     1,858,401  \\
    \multicolumn{1}{l}{Non-tertiary migrants } &                       8,159  &                     107,784  & 0     &                   10,800,000  \\
    \multicolumn{1}{l}{Population } &            43,700,000  &            130,000,000  &            1,331,475  &             1,240,000,000  \\
    \multicolumn{1}{l}{GDP p.c. PPP, destination country} &                    15,549  &                       13,547  &                        617  &                           58,227  \\
    \multicolumn{1}{l}{Distance in km between main cities} &                       7,411  &                         4,346  &                          55  &                           19,819  \\
    \multicolumn{1}{l}{Contiguity} &                         0.03  &                            0.16  & 0     & 1 \\
    \multicolumn{1}{l}{Common official or primary language} &                         0.11  &                            0.32  & 0     & 1 \\
    \multicolumn{1}{l}{Common colonizer post 1945} &                         0.05  &                            0.22  & 0     & 1 \\
    \multicolumn{1}{l}{Ever in colonial relationship} &                         0.01  &                            0.12  & 0     & 1 \\
    \multicolumn{1}{l}{Land area in square km (2020)} &              1,008,358  &                 2,461,710  & 718   &                   16,400,000  \\
    \midrule
    \end{tabular}
     \parbox{1\textwidth}{\footnotesize Observations: The number of observations (country-pairs) is 61,193 in Panel A and 25,189 in Panel B. The number of destination countries in Panel A is 146 in 2000, 147 in 2010, and 148 in 2020, while in Panel B, it is 91 in 2000 and 81 in 2010. The number of origin countries in Panel A is 148 in 2000, 149 in 2010, and 150 in 2020, whereas in Panel B, it is 147 in 2000 and 148 in 2010.}
\end{table}

\newpage

\section{Methods}\label{methods}
To intuitively understand our approach to measuring openness, imagine that two countries, A and B, are identical in many measurable respects: geographic location, population size, GDP per capita, land size, spoken languages, colonial history, and a number of other characteristics that determine the flows of people between countries. Based on these characteristics, we make a statistical prediction of the countries' immigration rate, and the estimate is 10 percent. However, we learn that the actual immigration rates of the two countries are 15 percent for country A and 5 percent for country B. Hence, country A has a 5 pp higher rate of immigration than expected, and country B has a 5 pp lower rate than expected. The difference between the actual and the expected immigration rates may be informative of their formal and informal norms and policies towards migrants. We call this difference \textit{residual immigration}. This residual immigration is estimated using gravity models of countries' bilateral migration and forms the basis for our measures of openness to migrants. 

The fundamental principle of the gravity model in economics is that larger economies tend to have a greater propensity to trade or experience migration flows, similar to how larger masses in physics exert a greater gravitational pull. These models generally posit that bilateral trade or migration between two countries is positively related to their economic sizes (usually measured by GDP) and inversely related to the distance between them, as distance increases the costs of trade \citep{anderson2011gravity}. 

In the migration literature, gravity models have been adapted to analyze and predict migration flows between regions or countries \citep{mayda2010international,ramos2016gravity,beine2016practitioners}. These models hypothesize that the migration flow between two locations is positively related to the population size of the source and destination (indicating more potential migrants and more opportunities at the destination, respectively) and negatively related to the distance between them, as greater distances often mean higher costs and more significant barriers to migration. Researchers have extended the model to include additional variables such as income levels, employment rates, political stability, or policy barriers, aiming to capture the complex decision-making process of potential migrants. The gravity model in this context helps in understanding patterns of human mobility, policy implications, and the socio-economic forces driving migration \citep{lewer2008gravity,poot2016gravity,bahar2018migration}.

We model the stock of migrants $M$ between country of origin $o$ and country of destination $d$ as a function of the distance between $o$ and $d$, their size in terms of population and land area, their level of income as captured in GDP per capita purchasing power parity (GDP p.c. PPP), cultural similarity as captured by shared official or primary language, shared history as captured by having had a colonial relationship, a dummy for having a shared border (contiguity), and the land size of the country of destination. Moreover, we'd like to control for the characteristics of the sending countries and changes in these characteristics (e.g., a conflict and its evolution). Equation \ref{eq: 1} specifies this. For our measures of openness, we are interested in the residual variation in $M_{od}$ as it pertains to the country of destination. Hence, we include country of origin fixed effects ($a_{o}$) and time fixed effects ($T$) interacted with the country of origin fixed effects ($a_{o} \times T$). This means that any variables which vary across countries of origin and not across countries of destination (e.g., the GDP p.c. PPP of countries of origin or incidence of conflicts) will be absorbed by these fixed effects. 

\begin{equation}\label{eq: 1}
\begin{split}
    M_{od}  = \beta_{1}log(pop_{d}) + \beta_{2}log(dist_{od}) + \beta_{3}log (GDPpc_{d}) \\
    + \beta_{4}language_{od} + \beta_{5}comcol_{od} + \beta_{6}coldepever_{od} \\
    + \beta_{7}contig_{od} + \beta_{8}log(land_{d})  + a_o + T + a_o \times T + \varepsilon_{od}
\end{split}
\end{equation}

$M_{od}$ is the stock of migrants from country of origin $o$ to country of destination $d$. In the bilateral matrix of migration stocks, about 75 percent of the cells are zero, meaning that 75 percent of the country pairs do not exchange any migrants. To estimate equation \ref{eq: 1}, we use a Poisson Pseudo Maximum Likelihood (PPML)  with high-dimensional fixed effects \citep{correia2020fast}. This model is well suited for modeling relationships where the outcome variable has many zeros and allows for the inclusion of various fixed effects. The residual $\varepsilon_{od}$ is of interest to us. This is the key input in the measurement of the openness to migration. 

\begin{table}[h]
    \centering
    \caption{Gravity Model Results, Main Specification.}
\begin{tabular}{lccc} \hline
 & (1) & (2) & (3) \\
Dependent var: & All migrants & Tertiary & Non-tertiary \\ \hline
 &  &  &  \\
$log(pop_{d})$  & 0.525*** & 0.556*** & 0.678*** \\
 & (0.0558) & (0.0408) & (0.0628) \\
$log(GDPpc_{d})$ & 1.359*** & 1.440*** & 1.216*** \\
 & (0.136) & (0.180) & (0.130) \\
$log(dist_{od})$ & -1.041*** & -0.595*** & -1.025*** \\
 & (0.107) & (0.111) & (0.122) \\
$contig_{od}$ & 1.188*** & 0.961*** & 1.502*** \\
 & (0.212) & (0.259) & (0.247) \\
$language_{od}$ & 1.080*** & 1.471*** & 1.111*** \\
 & (0.158) & (0.165) & (0.171) \\
$comcol_{od}$ & 1.251*** & 1.221*** & 1.376*** \\
 & (0.231) & (0.367) & (0.331) \\
$coldepever_{od}$ & 0.840*** & 0.953*** & 0.804*** \\
 & (0.189) & (0.177) & (0.215) \\
$log(land_{d})$ & 0.297*** & 0.356*** & 0.276*** \\
 & (0.0476) & (0.0576) & (0.0553) \\
Constant & -8.376*** & -15.28*** & -9.768*** \\
 & (1.734) & (2.027) & (1.823) \\
 &  &  &  \\
Observations & 61,193 & 25,189 & 25,189 \\
Pseudo R2 & 0.716 & 0.788 & 0.760 \\
 Wald chi2 & 804.8 & 1173 & 805.8 \\ \hline
\end{tabular}
\parbox{0.65\textwidth}{\footnotesize Note: Standard errors clustered by country of origin in parentheses. 
\\
*** p$<$0.01, ** p$<$0.05, * p$<$0.1.}
\end{table}


\section*{Measures of \textit{de facto} Openness to Migration}\label{sec:measures}

We consider two sets of residual-based measures of immigration openness. One uses the \textit{sum of residuals} to define openness, and the other uses the \textit{number} of positive residuals. We refer to the first as \emph{scale-based measures} and to the second as \emph{diversity-based measures}, as the number of positive residuals points to the diversity of places that a host country attracts people from.

\paragraph{Scale-based Measures}  We define openness to immigration of host country $d$, ($S_{d}$) as the ratio between the sum of all bilateral residuals by a country of destination ($res_{od}$) and the population of that country of destination ($pop_d$):

\begin{equation}
        res_{od} = I_{od}-\hat{I}_{od}
\end{equation}

\begin{equation}
        S_{d} = \frac{\sum^{148}_{o=1}res_{od}}{pop_d}
\end{equation}

Where $I_{od}$ is the migration stock and $\hat{I}_{od}$ is the predicted migration stock.

Such a measure rewards countries that host a large number of immigrants relative to the host country's population, independent of whether the immigrants come from a few or from many countries. Negative residuals are part of the sum as well. This measure is sensitive to single bilateral outliers. Such outliers sometimes present true acts of openness - such as Colombia opening up for migrants from Venezuela in 2015 (see \cite{bahar2021give} for details). Very often, however, the outliers are of a different nature. First, income imbalances between neighboring countries incentivize people to work and reside in two different places. For instance, many Slovaks and Hungarians work in Austria but prefer to reside in their respective countries of origin, where the cost of living is lower. In our estimates, the residual between Austria as a host country and Slovakia as a country of origin is negative, even though Austria is not restricting Slovaks from working and migrating there, understating Austria's openness to migrants.\footnote{Similar is the situation with many other European countries: Brits working in Ireland, Italian and Frenchmen working in Switzerland, Germans working in the Netherlands, and Albanians and North Macedonians working in Greece.} The second type of bilateral outlier captures situations in which a host country is unable to control its border, although it would like to. This is the case with the U.S.-Mexico  or the Pakistan-Afghanistan borders. Here, a sum-of-residuals measure may greatly overstate a country's openness to immigration. The third and last type of outlier is one where a single bilateral lack of cooperation overshadows a country's otherwise open stance on immigration. The Nordic countries, for instance, have a record of being open to refugees and economic migrants from various parts of the world. They are, however, relatively close to nationals from Russia. A single large negative residual with Russia overshadows their otherwise open social norms and policies towards migrants.
It is not possible to always correct this measure without making arbitrary choices about when an outlier is a true outlier and when it represents an act of openness or restrictiveness.\footnote{To mitigate some of the impact of single outlier residuals with neighboring countries, we also calculated these measures after excluding all residuals with contiguous countries. This correction helps in some cases, but in others, it makes the problem worse by punishing countries' genuine openness towards neighbors in trouble. Also, as demonstrated later in this section, neither of the two versions of scale-based openness performs as expected when correlated with \textit{de jure} measures and survey-based migrant acceptance measures (see Table \ref{tab:corr}).} Therefore, we consider alternative measures which do not suffer these shortcomings.

\paragraph{Diversity-based Measures.} Among several possibilities,
a simple count of the number of positive bilateral residuals by the host country shows a number of attractive features. First, compared to scale-based measures, the diversity-based measure is robust to single residual outliers. Second, compared to possible alternative diversity-based measures (e.g., the effective number of positive residuals), large bilateral events (e.g., Colombia's opening to Venezuelan migrants) correctly translate into a change in openness, the relative size of the other groups does not affect the overall measures. We tested the sensitivity of the measures to the cutoff point that defines when a residual is counted as a positive. We tried a range of cutoffs, e.g., 1 "excess" immigrant per 1 million host-country inhabitants, 5 per 1 million, 10 per 1 million, etc. We find that the measures become less stable when the cutoff is very low (e.g., 1 per million), but they are robust when using more conservative cutoffs (in the range of 5 to 10 per million).

\paragraph{Choosing Between Measures.} The two sets of measures rank countries differently on their openness, making it important to explain why we prefer one set of measures over the other. While both approaches have merit, the diversity-based measures capture our idea of openness better. Intuitively, they measure the degree to which a place has received people from places that are geographically, economically, and culturally different from themselves. Moreover, as discussed above, compared to the scale-based measures, bilateral outliers have a limited impact when using a diversity-based approach. Lastly, we check how the two types of measures correlate with existing \textit{de jure} measures of openness \citep{DEMIG2015} and migrant integration \citep{solano2020}. We also check how they correlate with Gallup Poll's Migrant Acceptance Index (MAI) \citep{fleming2018data}. 

We find that countries that are closed, as measured by their bilateral visa regimes, tend to be closed based on our diversity-based, but open based on our scale-based measures (variables $Visa_{do}$ and $Visa_{od}$ in Table \ref{tab:corr}. Moreover, while the scale-based measures are uncorrelated with Gallup Poll's MAI, countries that show higher MAI also tend to be more open in terms of diversity. Lastly, the Migration Integration Policy Index (MIPEX) is available for 56 of the 148 in our dataset, and for the years 2010 and 2020 \citep{MIPEX}. For this subset of countries, we find a strong positive correlation between MIPEX and diversity-based measures and a weaker negative one with scale-based measures (Table \ref{tab:corr}).

Overall, we prefer the diversity-based measures because they are aligned with our conceptual understanding of openness, they are robust to outliers, correlate strongly and in the expected direction with \textit{de jure} measures of openness, and a measure of immigrant acceptance.

\begin{table}[h]
  \centering
  \caption{Correlations between our Measures of Openness and Other Measures of Openness to Immigration \label{tab:corr}}
	\begin{tabular}{lcccccccc}
    \hline
     & (1) & (2) & (3) & (4) & (5) & (6) & (7) & (8) \\
    \hline
    (1) Diversity-based & 1 &       &       &       &       &       &       &  \\
    (2) Scale w. neighb. & 0.1453* & 1 &       &       &       &       &       &  \\
    (3) Scale wo. neighb. & 0.2442* & 0.8513* & 1 &       &       &       &       &  \\
    (4) $Visa_{do}$ & -0.1483* & 0.1479* & 0.0815 & 1 &       &       &       &  \\
    (5) $Visa_{od}$ & -0.5325* & 0.2114* & 0.0681 & 0.6395* & 1 &       &       &  \\
    (6) MAI (Rank) & -0.3630* & 0.0258 & -0.0558 & 0.0364 & 0.1777* & 1 &       &  \\
    (7) MAI & 0.3268* & -0.0135 & 0.0693 & -0.0229 & -0.1373* & -0.9860* & 1 &  \\
    (8) MIPEX Points & 0.5119* & -0.2718* & -0.1369 & -0.4414* & -0.6704* & -0.6293* & 0.5915* & 1 \\
    \hline
    \end{tabular}
\parbox{1\textwidth}{\footnotesize

  Note: $Visa_{do}$ counts the number of countries $o$ towards which host country $d$ imposes visa, while $Visa_{od}$ counts the number of countries $o$ which impose visas on $d$. MAI is derived from three questions about migrant acceptance from the Gallup World Poll 2016/2017 \citep{fleming2018data}. In MAI Rank, the most accepting country has rank 1. MIPEX Points evaluates 56 countries on 8 integration policy areas \citep{solano2020} The index used here is the composite one, reflecting all 8 areas.

  * Significant at 5\% level or better. Number of observations in correlations with openness measures: 435 ($Visa_{do}$), 440 ($Visa_{od}$), 387 (MAI), 104 (MIPEX).}
\end{table}%

\paragraph{Bilateral Examples.} To illustrate the approach's main components - the bilateral residuals from equation \ref{eq: 1} - we start by showing a few examples of open and closed countries. In the most open place in the world in 2020 - the United Kingdom - about 14 percent of the population was foreign-born, but a large part of this population came from places that are geographically, culturally, and economically different from the UK. We counted 62 such places using the default measure. In contrast, in the most open Gulf country - Kuwait - 68 percent of the population is foreign-born, but they came from 20 such places.

\begin{table}[h]
  \centering
  \caption{Examples of Open and Closed Places \label{tab:resid}}
    \begin{tabular}{p{6.045em}ccccccc}

    \hline
    \multicolumn{2}{c}{GBR} & \multicolumn{2}{c}{KWT} & \multicolumn{2}{c}{IND} & \multicolumn{2}{c}{COL} \\  \cmidrule(rl){1-2}  \cmidrule(rl){3-4}  \cmidrule(rl){5-6}     \cmidrule(rl){7-8} 

    Origin & \multicolumn{1}{p{3em}}{Share of host population} & \multicolumn{1}{p{6.045em}}{Origin} & \multicolumn{1}{p{3em}}{Share of host population} & \multicolumn{1}{p{6.045em}}{Origin} & \multicolumn{1}{p{3em}}{Share of host population} & \multicolumn{1}{p{6.045em}}{Origin} & \multicolumn{1}{p{3em}}{Share of host population} \\
    \midrule
    \multicolumn{1}{c}{POL} & 1.15\% & IND   & 23.97\% & NPL   & 0.01\% & VEN   & 1.19\% \\
    \multicolumn{1}{c}{PAK} & 0.57\% & EGY   & 8.74\% &       &       &       &  \\
    \multicolumn{1}{c}{ROU} & 0.47\% & BGD   & 7.80\% &       &       &       &  \\
    \multicolumn{1}{c}{DEU} & 0.34\% & PAK   & 7.10\% &       &       &       &  \\
    \multicolumn{1}{c}{LTU} & 0.27\% & PHL   & 4.38\% &       &       &       &  \\
    \multicolumn{1}{c}{ITA} & 0.25\% & IDN   & 2.25\% &       &       &       &  \\
    \multicolumn{1}{c}{NGA} & 0.21\% & YEM   & 0.90\% &       &       &       &  \\
    \multicolumn{1}{c}{USA} & 0.21\% & JOR   & 0.89\% &       &       &       &  \\
    \multicolumn{1}{c}{BGD} & 0.18\% & LKA   & 0.50\% &       &       &       &  \\
    \multicolumn{1}{c}{CHN} & 0.18\% & NPL   & 0.44\% &       &       &       &  \\
    \multicolumn{1}{c}{ZWE} & 0.17\% & ARE   & 0.44\% &       &       &       &  \\
    \multicolumn{1}{c}{KEN} & 0.16\% & SDN   & 0.23\% &       &       &       &  \\
    \multicolumn{1}{c}{AUS} & 0.16\% & GBR   & 0.13\% &       &       &       &  \\
    \multicolumn{1}{c}{ESP} & 0.15\% & USA   & 0.10\% &       &       &       &  \\
    \multicolumn{1}{c}{CAN} & 0.14\% &       &       &       &       &       &  \\
    \multicolumn{1}{c}{HKG} & 0.13\% &       &       &       &       &       &  \\
    \multicolumn{1}{c}{JAM} & 0.13\% &       &       &       &       &       &  \\
    \multicolumn{1}{c}{LVA} & 0.12\% &       &       &       &       &       &  \\
    \multicolumn{1}{c}{LKA} & 0.12\% &       &       &       &       &       &  \\
    \multicolumn{1}{c}{BGR} & 0.12\% &       &       &       &       &       &  \\
    \multicolumn{1}{c}{SOM} & 0.11\% &       &       &       &       &       &  \\
    \multicolumn{1}{c}{PRT} & 0.11\% &       &       &       &       &       &  \\
    \multicolumn{1}{c}{IND} & 0.11\% &       &       &       &       &       &  \\
    \multicolumn{1}{c}{MYS} & 0.10\% &       &       &       &       &       &  \\
    \multicolumn{1}{c}{SVK} & 0.10\% &       &       &       &       &       &  \\
    \midrule
    Foreigners in total population & 12.9\% &       & 68.4\% &       & 0.33\% &       & 3.7\% \\
    \bottomrule
    \end{tabular}%
\end{table}%

In closed places, which are the majority of territories in the World, these numbers are close to zero. In India, for instance, similar to other parts of South Asia, only 0.33 percent of the 2020 population was foreign-born. Within this population, India was open towards one place according to our measure - Nepal\footnote{We are unlikely to underestimate India's openness. Despite its sheer population size, the size of the foreign population in India is very small. After the Nepali population of about 734,000, the largest populations came from the UAE (38,000) and the USA, (33,000).}. Similarly, according to the 2020 estimates, Colombia was open to two places - Venezuela and the USA. Prior to its opening to refugees fleeing the Venezuelan crisis, Colombia was only open to the USA, and less than 0.3 percent of its population was foreign-born. Table \ref{tab:resid} shows the countries towards which the UK, Kuwait, India, and Colombia had positive residuals, with a cutoff of 100 per million, with the exception of India, where we had to apply the 10 per million cutoff in order to detect openness\footnote{Our default cutoff is 10 per million, but for illustration, we chose a more conservative cutoff here.}.

\section{Findings from our Measures of Openness to Immigration}\label{findings}

\begin{figure}[h]
    \centering
    \caption{Diversity-based Openness in 2020}
    \includegraphics[width=1\linewidth]{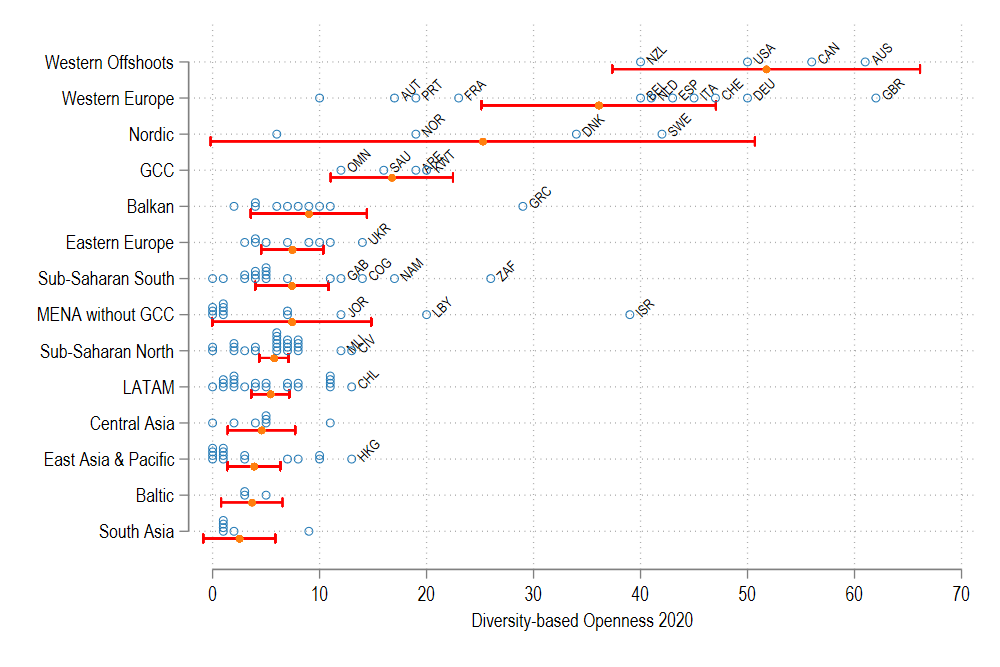}
    \label{fig:open2020}
    Note: ISO3 labels are shown for places that have an openness score of 15 or higher.
\end{figure}

Today, most places in the World are rather closed to immigration, and few places are extremely open (Figure \ref{fig:open2020}). This resonates with previous findings about the Global patterns of immigration for earlier decades \citep{czaika2014globalization}. Broadly speaking, the exceptions to the World's lack of openness are the Western Offshoots, much of Western Europe, the Nordic countries except for Finland, and the Gulf countries. In other regions, a few other countries are very open - Israel\footnote{Israel is an exception of a different kind. The 1950 Law or Return grants that every Jew, regardless of their place of birth, can migrate to Israel and gain Israeli citizenship. Non-Jewish people can migrate and become citizens if they are in a family union with a Jewish person.}, Greece, Libya, and South Africa, for example.

Within more closed regions, the following places are moderately open: Namibia (open to 17 places), the Republic of Congo (14), Gabon (12) and Mozambique (11) in the Sub-Saharan South; Côte d'Ivoire (13) and Mali (12) in the Sub-Saharan North; Ukraine (14) and Russia (11) in Eastern Europe; Chile (13), Peru (11), Costa Rica (11), Panama (11), and Ecuador (11) in Latin America; Kyrgyzstan (11) in Central Asia; Turkey (11) in the Balkan; and Hong Kong (13), Singapore (10) and South Korea (10) in East Asia and Pacific. The most open place in South Asia is Bangladesh (9 places). Among the three Baltic states, Estonia is the most open one (5 places).

Places that are open tend to be open to people with different levels of education. The coefficient of correlation between openness towards tertiary and non-tertiary educated immigrants is 0.827. This is probably due to the fact that our measures reflect long-run tendencies of countries instead of short-term migratory flows, and over time, with diaspora network formation \citep{mckenzie2010self,prieto2024diaspora} playing an important role, the inflows of less and more skilled migrants become tightly connected. In 2000, among the 91 countries for which we have data on the educational level of the immigrants, on average, countries were open to 6.7 places for the tertiary educated migrants and to 10.1 for the non-tertiary educated. In 2010, for 81 available countries, they were open on average to 8.2 and 12.2 places respectively.

\begin{figure}[h]
    \caption{Openness towards more and less educated immigrants}
    \includegraphics[width=1\linewidth]{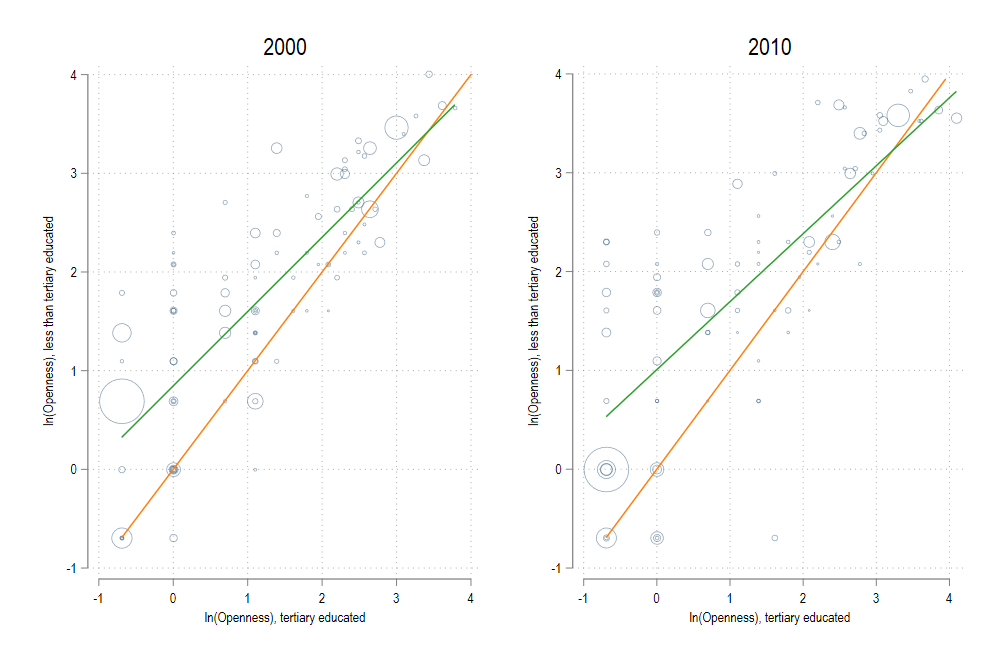}
    \label{fig:open_tnt}
    Note: The size of the circle is proportional to the population size of a place. Zero values were replaced with 0.5 to avoid losing fully closed places in the log transformation of openness. The orange line is the 45-degree line. The green shows the linear fit between the two variables.
\end{figure}

This bias towards less-educated migrants is visible in Figure \ref{fig:open_tnt}. The figure shows the log-transformed measures of openness\footnote{Zero values were replaced with 0.5 to avoid losing fully closed places from the figure.} for the tertiary-educated immigrants (x-axis) and less than tertiary-educated ones (y-axis). The orange line is a 45-degree line, and the green shows the linear fit between the two variables. In the positioning of most countries above the 45-degree line, we see that most countries are more open towards less educated migrants than towards highly educated ones.

\begin{figure}[H]
    \centering
    \caption{Change in Openness 2000-2010 and 2010-2020}
    \includegraphics[width=1\linewidth]{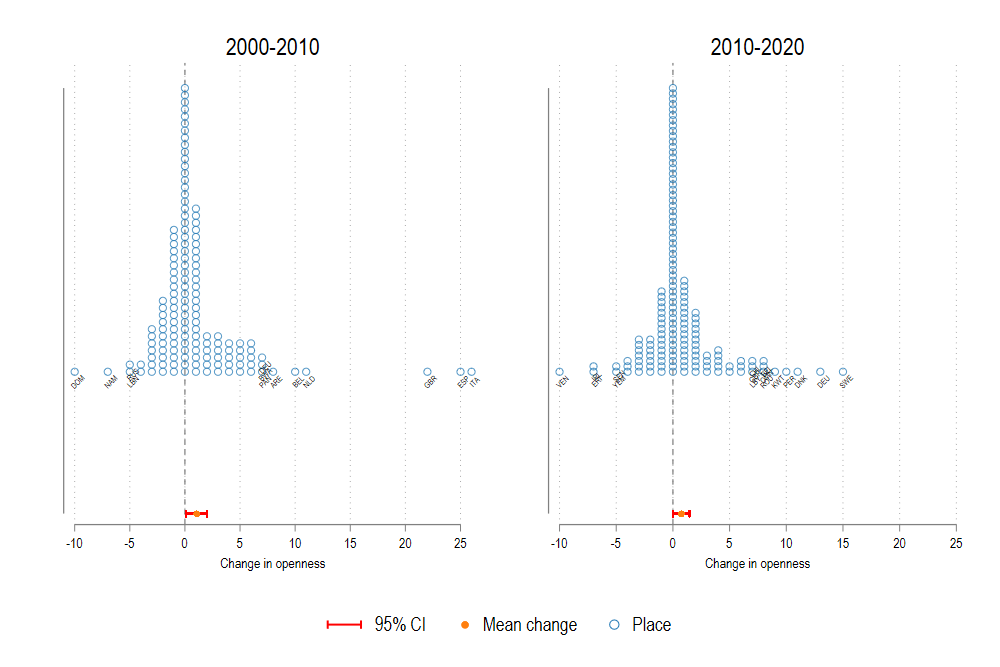}
    \label{fig:open_chg}
\end{figure}

Between 2000 and 2010, on average, economies around the world became more open to one other place (the average change in openness was 1.05), while between 2010 and 2020, on average, this change was 0.73 (Figure \ref{fig:open_chg}). However, the positive changes in openness were driven by a few regions and places that experienced staggering increases in openness, such as Italy, Spain, and the U.K. between 2000 and 2010, and Germany and Sweden between 2010 and 2020.

Looking by region, the regions that opened the most in the decade 2000-2010 were Western Europe, the Western Offshoots,  the Nordic countries, GCC, and to a much lesser extent, South Asia and the Balkan, while Central Asia and the Baltic states became slightly less open.

\begin{figure}[H]]
    \centering
    \caption{Change in Openness by region and decade}
    \includegraphics[width=1\linewidth]{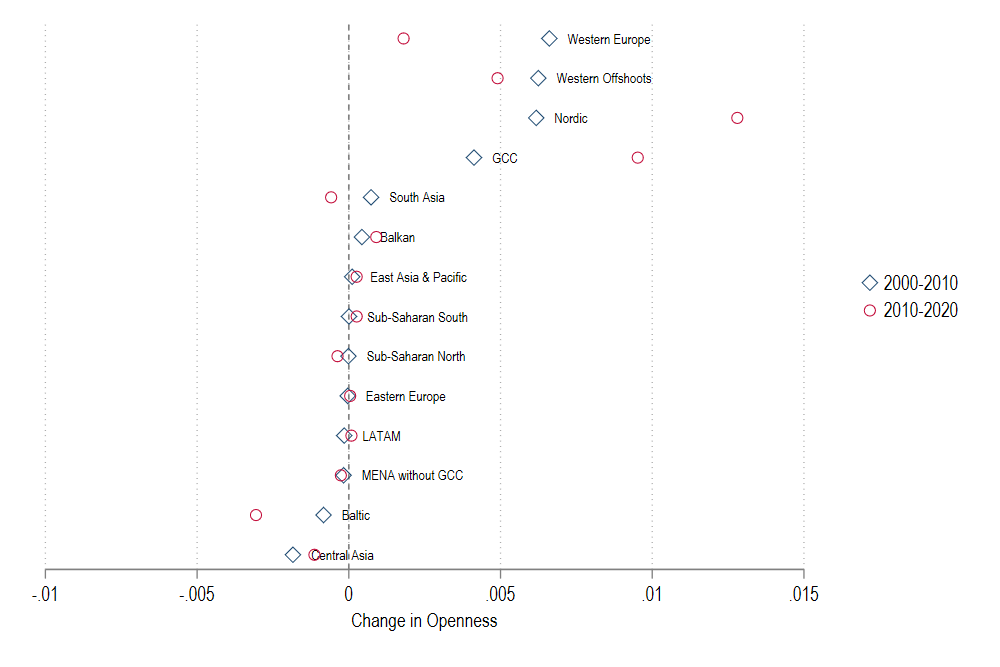}
    \label{fig:open_chg_reg}
\end{figure}

\newpage

\section{Openness, Aging, and Wages}\label{validation}

In this section, we test two hypotheses about the impact of openness on country-level outcomes. We study what happens to demographic change and real wages when countries become more open. The results of this analysis serve as a further validation of our measures. They show that our measures are predictive of country-level outcomes and that these relationships behave according to expectations.

It has been argued that immigration is a potent lever for improving a country's fiscal sustainability \citep{storesletten2000sustaining, kerr2011economic, dustmann2014fiscal}. Economic migrants tend to be younger and more economically active than the non-migrant population. Immigration could, therefore, help aging economies counteract the fiscal burden of retirement by changing a country's demographic pyramid. We can capture this aspect of demographic change by looking at the old-age dependency ratios. This is the ratio of the elderly (65+) and the working-age population (15-64). We expect that countries that become more open over time see slower growth in their old-age dependency ratio. 

Moreover, policymakers often promote immigration as a solution to labor and skill shortages. \cite{Teivainen2021, McGuinness2023, Lutz2024} present contemporary examples. Guest worker programs such as the H-2A and H-2B visa programs in the United States, Canada's  Temporary Foreign Worker Program, or the Kafala system in the GCC countries, are some better known examples. See \cite{castles1986guest, castles2006guestworkers} for a review of guest worker programs. Such programs can facilitate the availability of (foreign) labor and skills in the hosting countries. As the labor supply constraint is relaxed, the upward pressure on domestic wages should decline. This is what the canonical economic model of the impact of immigration on the labor market \citep{altonji1991effects, borjas2014immigration} predicts: within similar skill groups, increased labor supply should temporarily decrease the wages of the existing labor force and increase the marginal productivity of capital. While the empirical tests of the above have been mixed, \cite{dustmann2016impact} have helped settle important sources of divergence in empirical findings. They identify, what they name a 'pure spatial approach', to estimate the right kind of structural parameter for policy design - the \textit{total} wage effect of immigration. Within this approach, most studies find that the adverse effects of immigration are concentrated among the low-skilled, while the effects on the skilled labor force tend to be positive. For instance, for the United States, \cite{monras2020immigration} finds a lasting state of deteriorated labor market for low-skilled natives who entered the labor force years of high-immigration. For the UK, \cite{dustmann2013effect} find mainly zero or small negative effects of immigration on the wages of the low-skilled and mainly positive effects on the wages of the highly-skilled, and for Germany, \cite{dustmann2015industries} find a negative average impact of immigration on wages, results mainly driven by jobs in the non-tradable sector, and by workers with low and medium-level skills.  Hence, from the canonical model, we would expect a zero or a negative correlation between increases in openness and wage growth. Moreover, we expect the average effects to be driven by the less-skilled workers. Although we cannot test this directly because we do not distinguish between the wages of the less- and the better-skilled workers, it is reasonable to assume that the average is largely driven by earners with lower levels of skills:  non-tertiary workers tend to dominate the domestic labor force, earlier in the study, we showed that countries tend to be more open to non-tertiary migrants. 

To test the effect of openness to immigration on the changes in age, we estimate two first difference equations:

\begin{equation}
    \begin{split}
    \Delta Old_{dt} = \beta_0 + \beta_1 \Delta Open_{dt} + \beta_2 Open_{d,t-10} \\
    + \beta_3 Old_{d,t-10} + \beta_4 ln(GDPpc)_{d,t-10} + T_t + \varepsilon_{dt}      
    \end{split}
\end{equation}

\begin{equation}
    \begin{split}
    \Delta ln(w)_{dt} = \beta_0 + \beta_1 \Delta Open_{dt} + \beta_2 Open_{d,t-10} \\
    + \beta_3 ln(w)_{d,t-10} + \beta_4 Old_{d,t-10} + T_t + \epsilon_{dt}
    \end{split}
\end{equation}

Here, $\Delta$ are ten-year periods 2000-2010 and 2010-2020. Among the outcome variables,  $Old$ is the old-age dependency ratio, and $ln(w)$ is the natural log of wages, defined as a real GDP p.c. per employee. $Open$ is our default measures of diversity-based openness, $ln(GPDpc)$ is the natural log of the GDP p.c.. This should capture various factors related to the country's wealth and level of development that might also affect the country's demographic transition, $T$ are time-fixed effects, basically a dummy for 2020, and $\varepsilon$ and $\epsilon$ are the residuals. The subscripts $d$ and $t$ refer to the country of the immigrant destination and time.

We find that open countries age slower, and that further opening is related to a slowdown of the demographic transition. We also find that open countries experience slower real wage growth, and that further opening is related to a its further slowdown.

More precisely, a greater openness of one place at the beginning of a decade is associated with a slower decennial growth of the dependency ratio of 0.054 percentage points. Moreover, countries that opened to one more place within a ten-year period, experienced an additional slowdown of 0.105 pp (Column 3, Table \ref{tab:oadr}). The control variables correctly show that aging has been accelerating globally (positive coefficient of the lagged dependent variable) and that richer countries are aging faster. The openness variables significantly improve the fit of the model. Comparing columns 1 and 3, Table \ref{tab:oadr}, we see that the adjusted R-squared increased from 0.581 to 0.636.

We also find negative partial correlations between openness and wage growth. Being open to one more place at the beginning of a decade correlates with 0.27 percent lower real wage growth in the following decade, and increasing openness by one place correlates with 1.24 percent lower decennial wage growth (Column 3, Table \ref{tab:lnw}). The controls also have the expected signs: higher initial wages predict slower subsequent growth, and higher old-age dependency ratios predict more wage growth. We interpret this to mean that tighter labor markets (fewer working-age individuals) lead to higher wage growth. The added variables of interest significantly improve the fit of the model. Comparing columns 1 and 3 of Table \ref{tab:lnw}, we see that the adjusted R-squared improved from 0.197 to 0.243 after adding the openness variables. 


\begin{table}[h]
    \centering
    \caption{Openness and Aging\label{tab:oadr}}
\begin{tabular}{lccc} \hline
 & (1) & (2) & (3) \\ \hline
 &  &  &  \\
$\Delta Open$ &  &  & -0.105*** \\
 &  &  & (0.0244) \\
$Open_{t-10}$ &  & -0.0583*** & -0.0543*** \\
 &  & (0.0134) & (0.0128) \\
$Old_{t-10}$ & 0.183*** & 0.198*** & 0.203*** \\
 & (0.0212) & (0.0202) & (0.0200) \\
$ln(GDPpc)_{t-10}$ & 0.374*** & 0.545*** & 0.641*** \\
 & (0.0915) & (0.100) & (0.102) \\
2020 dummy & 1.557*** & 1.560*** & 1.493*** \\
 & (0.204) & (0.194) & (0.188) \\
Constant & -4.163*** & -5.223*** & -5.978*** \\
 & (0.627) & (0.695) & (0.719) \\
 &  &  &  \\
Observations & 286 & 286 & 286 \\
R-squared & 0.585 & 0.623 & 0.642 \\
 Adj. R2 & 0.581 & 0.618 & 0.636 \\ \hline
\multicolumn{4}{c}{ Outcome variable: $\Delta$ old-age dependency ratio} \\
\multicolumn{4}{c}{ Robust standard errors in parentheses} \\
\multicolumn{4}{c}{ Significant at: *** p$<$0.01, ** p$<$0.05, * p$<$0.1} \\
\end{tabular}
\end{table}

\begin{table}[h]
    \centering
    \caption{Openness and Wage Growth \label{tab:lnw}}
    \begin{tabular}{lccc} \hline
     & (1) & (2) & (3) \\ \hline
     &  &  &  \\
    $\Delta Open$ &  &  & -0.0124*** \\
     &  &  & (0.00285) \\
    $Open_{t-10}$ &  & -0.00336*** & -0.00267*** \\
     &  & (0.000926) & (0.000938) \\
    $ln(w)_{t-10}$ & -0.102*** & -0.0937*** & -0.0840*** \\
     & (0.0173) & (0.0176) & (0.0165) \\
    $Old_{t-10}$ & 0.00832*** & 0.00974*** & 0.0107*** \\
     & (0.00222) & (0.00232) & (0.00232) \\
    2020 dummy & -0.0881*** & -0.0877*** & -0.0943*** \\
     & (0.0251) & (0.0249) & (0.0246) \\
    Constant & 1.143*** & 1.076*** & 0.977*** \\
     & (0.160) & (0.161) & (0.153) \\
     &  &  &  \\
    Observations & 281 & 281 & 281 \\
    R-squared & 0.206 & 0.223 & 0.256 \\
     Adj. R2 & 0.197 & 0.212 & 0.243 \\ \hline
    \multicolumn{4}{c}{ Outcome variable: $\Delta ln(w)$} \\
    \multicolumn{4}{c}{ Robust standard errors in parentheses} \\
    \multicolumn{4}{c}{ *** p$<$0.01, ** p$<$0.05, * p$<$0.1} \\
    \end{tabular}
\end{table}

\newpage
\section{Conclusion}\label{conclusions}
We tend to think of countries' openness to immigration as a function of governments' policies and administration. It is much more than that. Here, we put forward a way of measuring openness to immigration. We propose country-specific measures of \textit{de facto} openness to immigration and perform a number of exercises to demonstrate their construct validity. The measures complement existing \textit{de jure} measures of immigration but help contrast policy intentions with immigration outcomes. Using the measures, we show that the World became more open in the first decade of the millennium, but less so in the second. With some exceptions, the opening was greatly driven by the Western World and the Gulf countries. We also show that countries that are open for high-skilled migration tend to be open for low-skilled migration as well and that the openness towards low-skilled migration is greater than the openness towards high-skilled migration, with only few exceptions. Lastly, we observe that increases in country-level openness in the last two decades, other factors equal were accompanied by a slowdown of countries' demographic transitions, as well as a slower real wage growth, arguably because openness helps places close labor and skill shortages.   

The proposed measures can be used to contrast long-run trends in immigration policy with changes in openness since the start of the millennium. Furthermore, our contribution presents a path towards studying the impact of openness on outcomes such as innovation, production diversification, entrepreneurship, and voting behavior, among other outcomes, but also can help us understand why some societies open over time while most do not.  

Our approach also has limitations. \textit{De facto} Openness is better suited to capture long-run trends of countries, and it is less suited to capture short-run fluctuations in openness. It cannot be reliably estimated for small territories, which is why we focus on places with at least 1.2 million in population. It is based on counting nationalities as opposed to ethnic groups, which can overstate openness in some cases and underestimate it in others. Lastly, although we control for country-of-origin fixed effects in the gravity models, we are unable to control for the differential selection of same-place-of-origin migrants into different host countries. While \textit{de facto} openness mainly captures the collective long-run efforts of a place to open for immigration, it also partially incorporates the degree to which migrants choose their host destination.

\backmatter


\section*{Acknowledgements}
We thank Lucila Venturi Grosso and Ricardo Villasmil for valuable inputs and insights, Sarah Bui, Alice Zhang, Margarita Isaacs, and Zahra Asghar for outstanding research assistance. This research was funded by the Templeton World Charity Foundation, Inc. through a grant (Grant number TWCF-2022-30478, “Leveraging the Global Talent Pool to Jumpstart Prosperity in Emerging Economies”, granted to the Growth Lab at Harvard University).

\bibliography{sn-bibliography}

\begin{appendices}

\section{Supplementary Materials}
We encourage the readers to explore the interactive \href{https://vis.csh.ac.at/leveraging-global-talent/}{website} featuring the measures of openness. The created \href{https://github.com/LjubicaN/Openness}{openness measures} are available to the public.

\section{Gravity Models: Choosing among Variety of Models}\label{A1}
Before converging on the model presented in equation \ref{eq: 1}, we experimented with various model specifications. For instance, in addition to the above-included variables, we have estimated models that include other explanatory variables of immigration such as the size of the existing diaspora of country $o$ in country $d$, the similarity of the product portfolio across country pairs, and shared religion. We also estimated models without colonial controls and without land area controls. Moreover, we specified models with constant GDP p.c. and GDP p.c. PPP. We also specified models of migration flows vs. migration stocks. Lastly, we estimated variants of the models that include controls for the size of the existing diasporas of sending countries.   

We used three criteria to choose the best model. First, migration flows are significantly noisier than migration stocks, and this is the main reason why we decided to work with migration stocks as the outcome variable. Second, we studied how sensitive the distribution of the residuals is to the inclusion or exclusion of a variable. Third, we looked at general measures of model fit such as the pseudo R-squared. 

The distribution of the residuals was not very sensitive to the inclusion or exclusion of variables such as religion and colonial ties, but the inclusion of colonial ties improved the model fit, while religion didn't. Moreover, while the values of individual countries were sometimes significantly affected, the overall ranking of these countries using our preferred measures of openness did not change.

Including land area improved the ranking of densely populated places (e.g., Hong Kong and a number of European countries), and worsened the ranking of sparsely populated countries (e.g., Sweden, USA, Canada, Australia). The inclusion of diaspora controls made a difference for a number of countries, but since it requires the use of lagged values, it came at a price - the loss of one decade of data. Since the inclusion of diaspora controls didn't significantly change the rank of the affected countries, we opted for a model that does not include the diaspora size as a control.

Lastly, we intentionally do not control for \textit{de jure} measures of immigration, as we'd like to ensure that the variation in the level of immigration that is due to legal migration measures and official migration policy remains in the residual. 



\end{appendices}


\end{document}